\documentclass[preprint,12pt, a4paper]{elsarticle}

\usepackage{amssymb}
\usepackage{lineno}

\usepackage{float}
\usepackage{subfig}

\usepackage{mathtools}

\usepackage{listings}
\usepackage{xcolor}
\definecolor{codegreen}{rgb}{0,0.6,0}
\definecolor{codegray}{rgb}{0.5,0.5,0.5}
\definecolor{codepurple}{rgb}{0.58,0,0.82}
\definecolor{backcolour}{rgb}{0.95,0.95,0.92}

\lstdefinestyle{mystyle}{
    backgroundcolor=\color{backcolour},   
    commentstyle=\color{codegreen},
    keywordstyle=\color{magenta},
    numberstyle=\tiny\color{codegray},
    stringstyle=\color{codepurple},
    basicstyle=\sffamily\scriptsize,
    breakatwhitespace=false,         
    breaklines=true,                 
    captionpos=b,                    
    keepspaces=false,                 
    numbers=none,                    
    numbersep=2pt,                  
    showspaces=false,                
    showstringspaces=false,
    showtabs=false,                  
    tabsize=2
}

\usepackage{hyperref}

\restylefloat{table}

\journal{SoftwareX}

\begin{document}

\begin{frontmatter}

%% Title, authors and addresses

%% use the tnoteref command within \title for footnotes;
%% use the tnotetext command for theassociated footnote;
%% use the fnref command within \author or \address for footnotes;
%% use the fntext command for theassociated footnote;
%% use the corref command within \author for corresponding author footnotes;
%% use the cortext command for theassociated footnote;
%% use the ead command for the email address,
%% and the form \ead[url] for the home page:
%% \title{Title\tnoteref{label1}}
%% \tnotetext[label1]{}
%% \author{Name\corref{cor1}\fnref{label2}}
%% \ead{email address}
%% \ead[url]{home page}
%% \fntext[label2]{}
%% \cortext[cor1]{}
%% \address{Address\fnref{label3}}
%% \fntext[label3]{}

\title{NL4Py: Agent-Based Modeling in Python with Parallelizable NetLogo Workspaces}

%% use optional labels to link authors explicitly to addresses:
%% \author[label1,label2]{}
%% \address[label1]{}
%% \address[label2]{}

\author[1]{Chathika Gunaratne}
\author[1]{Ivan Garibay\corref{cor1}\fnref{label2}}

\address[1]{Complex Adaptive Systems Lab\\
Department of Industrial Engineering and Management Systems,\\  
College of Engineering and Computer Science,\\
University of Central Florida\\
Orlando, FL}
 \ead{igaribay@ucf.edu}
\begin{abstract}
%% Text of abstract 
%Ca. 100 words
External control of agent-based models is vital for complex adaptive systems research. Often these experiments require vast numbers of simulation runs and are computationally expensive.
NetLogo is the language of choice for most agent-based modelers but lacks direct API access through Python.
NL4Py is a Python package for the parallel execution of NetLogo simulations via Python, designed for speed, scalability, and simplicity of use. NL4Py provides access to the large number of open-source machine learning and analytics libraries of Python and enables convenient and efficient parallelization of NetLogo simulations with minimal coding expertise by domain scientists.
\end{abstract}

\begin{keyword}
%% keywords here, in the form: keyword \sep keyword
NL4Py \sep Python \sep NetLogo \sep agent-based modeling \sep complex adaptive systems \sep parameter calibration
%keyword 1 \sep keyword 2 \sep keyword 3

%% PACS codes here, in the form: \PACS code \sep code

%% MSC codes here, in the form: \MSC code \sep code
%% or \MSC[2008] code \sep code (2000 is the default)

\end{keyword}

\end{frontmatter}

\section*{Required Metadata}
\label{}

\section*{Current code version}
\label{}

%Ancillary data table required for subversion of the codebase. Kindly replace examples in right column with the correct information about your current code, and leave the left column as it is.

\begin{table}[H]
\begin{tabular}{|l|p{6.5cm}|p{6.5cm}|}
\hline
\textbf{Nr.} & \textbf{Code metadata description} & \textbf{Please fill in this column} \\
\hline
C1 & Current code version & 0.9.0 (commit 6c4c2a0) \\
\hline
C2 & Permanent link to code/repository used for this code version & \url{https://github.com/chathika/NL4Py} \\
\hline
C3 & Code Ocean compute capsule & Not applicable\\
\hline
C4 & Legal Code License   & GNU General Public License 3.0 (GPLv3) \\
\hline
C5 & Code versioning system used & git \\
\hline
C6 & Software code languages, tools, and services used & Python \\
\hline
C7 & Compilation requirements, operating environments \& dependencies & NetLogo; Java, Python 3.6+\\
\hline
C8 & If available Link to developer documentation/manual & \url{https://github.com/chathika/NL4Py/blob/master/Readme.md} \\
\hline
C9 & Support email for questions & chathika@knights.ucf.edu \\
\hline
\end{tabular}
%\caption{Code metadata (mandatory)}
\caption{Code metadata}
\label{} 
\end{table}

%\linenumber

%% main text

%The permanent link to code/repository or the zip archive should include the following requirements: 

%README.txt and LICENSE.txt.

%Source code in a src/ directory, not the root of the repository.

%Tag corresponding with the version of the software that is reviewed.

%Documentation in the repository in a docs/ directory, and/or READMEs, as appropriate.

\section{Motivation and significance}
\label{sec:motivation}

%Introduce the scientific background and the motivation for developing the software.
%Explain why the software is important, and describe the exact (scientific) problem(s) it solves.

%Indicate in what way the software has contributed (or how it will contribute in the future) to the process of scientific discovery; if available, this is to be supported by citing a research paper using the software.
%Provide a description of the experimental setting (how does the user use the software?).
%Introduce related work in literature (cite or list algorithms used, other software etc.).

Agent-based modeling is a widely used modeling and simulation technique where complex adaptive systems \cite{mitchell2009complexity,farmer2009economy,janssen2006empirically,axtell2008rise,siebers2010discrete} are replicated as the result of micro-scale interactions of autonomous agents. Agent-based models are capable of reproducing emergent properties of real-world complex adaptive systems such as dynamic equilibria, non-linearity, and high sensitivity to initial conditions. Agent-based models often produce a vast space of possible macro-scale outcomes \cite{lee2015complexities} that require rigorous statistical scrutiny to characterize.
Techniques used range from simple factorial experiments, Monte-Carlo experiments, single or multi-objective parameter calibration/optimization \cite{stonedahl2010behaviorsearch}, sensitivity analysis \cite{lee2015complexities,ligmann2014using}, model/rule discovery \cite{gunaratne2017alternate}, and pattern identification \cite{cherel2015beyond}. Performing such experiments requires controlling agent-based models through an external application, perferrable written in a scripting language, rich with quick and easy-to-use statistical, computational, and machine learning libraries that are easily parallelizable.

At the time of writing, NetLogo \cite{wilensky1999netlogo,wilensky2015introduction} is the most popular agent-based modeling toolkit used in computational research. NetLogo is a dynamically-typed, pseudo-functional programming language, influenced by Logo and Lisp, accompanied by a drag-and-drop graphical user interface builder for parameter controls and model output display, making it quick and easy to learn. For this reason, NetLogo has enabled domain scientists with modest, to no, programming skills to quickly develop and experiment with agent-based models. While many other agent-based modeling frameworks exist with varying popularity (eg. Repast Simphony/HPC \cite{north2013complex,collier2013parallel}, MASON \cite{luke2005mason}, AnyLogic \cite{Borshchev:2013:ANR:2675807.2675980}), NetLogo has remained the platform of choice for many conducting agent-based modeling research.
Python has quickly become the preferred language for machine learning and data analysis, thanks to its extensive, well-maintained collection of open-source machine learning and statistics libraries, such as Numpy \cite{oliphant2006guide}, SciPy \cite{mckinney-proc-scipy-2010}, Pandas \cite{mckinney2011pandas}, SALib \cite{herman2017salib}, Scikit-learn \cite{pedregosa2011scikit}, DEAP \cite{DEAP_JMLR2012}, and Matplotlib \cite{hunter2007matplotlib}, to name a few. Additionally, Python's ease of use and short learning curve paired with browser-based integrated development environments such as Jupyter notebook \cite{Kluyver:2016aa} has made it the language of choice for many looking to perform quick, shareable, data intensive experiments.

There has been an increasing need for researchers to have access to a NetLogo controlling library for Python for performing post-simulation, model analysis. However, NetLogo is written in a combination of Java and Scala, making it difficult to directly call and control NetLogo procedures through Python. Instead, external machine learning or data analysis code can access the NetLogo controlling application programming interface (API). This is straightforward through languages such as Java that interface directly with the NetLogo controlling API. Tools implemented in Java or Scala, such as NetLogo's BehaviorSpace, for factorial experiments, and BehaviorSearch \cite{stonedahl2010behaviorsearch}, for model calibration, integrate easily with the NetLogo controlling API. However, working with the NetLogo controlling API through Python requires the researcher to have the programming expertise necessary to work with the Java native interface (JNI) or remote procedure calls to the Java virtual machine (JVM) through sockets. Additionally, many statistical techniques used to analyze agent-based model output require vast counts of model runs under varying parameter configurations, often requiring the application to manage parallel NetLogo simulation execution.

NL4Py was developed to provide programmatic control of NetLogo models through Python and allow for parallel simulation. Similar to RNetLogo for R \cite{thiele2014r}, with NL4py, computational social scientists are enabled to utilize the many machine learning and data analytics libraries of Python to investigate the dynamics produced by their NetLogo models. NL4Py was initially developed to support such a framework, \textit{evolutionary model discovery}, which combines genetic programming and random forests for causal inference in complex adaptive systems \cite{gunaratne2021inferring,gunaratne2020evolutionary,gunaratne2019evolutionary}, and was later released as a general purpose library. NL4Py enabled evolutionary model discovery to easily interface with NetLogo models via Python and utilize the evolutionary algorithms library DEAP \cite{DEAP_JMLR2012}, the machine learning library SciKit-Learn \cite{pedregosa2011scikit}, and the scientific computing library SciPy \cite{mckinney-proc-scipy-2010}. Other such projects have also benefited from NL4Py towards the analysis of agent-based models in NetLogo \cite{elmenreich2021artificial, fullsack2020predicting, vandewalle2019integrating, pike2019standardizing, von2020modeling}. NL4Py was developed with the goals of usability, rapid parallel execution, and model parameter access in mind. We also compare NL4Py's performance with that of PyNetLogo \cite{jaxa2018pynetlogo}, an alternate solution for controlling NetLogo models in Python. Unlike PyNetLogo, which uses the Java native interface (JNI) framework to access the JVM, NL4Py, inspired by \cite{Py2NetLogo}, employs a client-server architecture via Py4J \cite{Py4J}, a Python-Java bridging package that uses socket communication.

\section{Software Description}
\label{sec:description}
NL4Py is a Python library that facilitates the external deployment, execution, and reporting of parallel simulations of NetLogo agent-based models. NL4Py supports Windows, MacOS, and Linux, and runs on Python 3. 

\subsection{Software Architecture}
\label{sec:architecture}

\begin{figure}[!h]
  \centering
  \includegraphics[width=\linewidth]{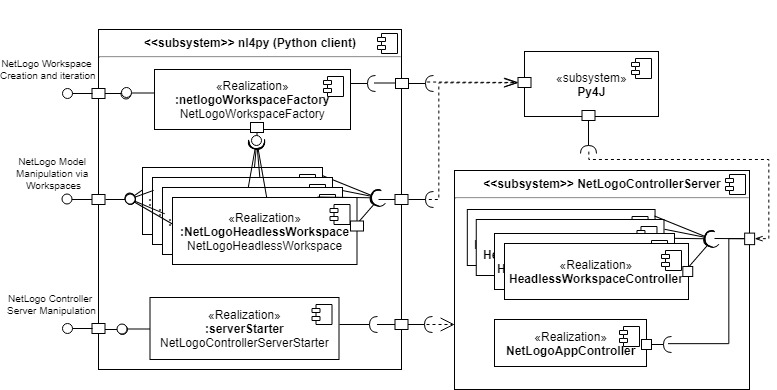}
  \caption{UML Component Diagram of NL4Py}
  \label{fig:componentdiagram}
\end{figure}

%Give a short overview of the overall software architecture; provide a pictorial component overview or similar (if possible). If necessary provide implementation details.
NL4Py uses a client-server architecture and consists of two main components, 1) the Python-based NL4Py client and 2) the Java-based \\ \lstinline[language=Python]{NetLogoControllerServer} JAR executable, as shown in Fig. \ref{fig:componentdiagram}. The client communicates with the \lstinline[language=Python]{NetLogoControllerServer} through sockets using the Py4J library. NetLogo headless workspaces can be accessed via the NetLogo controlling API for Java or Scala applications and is used by the \lstinline[language=Python]{NetLogoControllerServer} to create and manage NetLogo simulations upon request. The client-server architecture allows headless workspaces to be run in parallel as Java threads on the \lstinline[language=Python]{NetLogoControllerServer}, independent of the user's Python application code. This eliminates the need for users to have to manage the connection to the JVM, thread/process creation, and garbage collection of multiple headless workspaces from their Python application code.

The NL4Py Python client provides thread-safe \lstinline[language=Python]{NetLogoHeadlessWorkspace} objects to the Python application developer, created according to the factory design pattern. Each \lstinline[language=Python]{NetLogoHeadlessWorkspace} object is mapped to a \\ \lstinline[language=Python]{HeadlessWorkspaceController} object on the \lstinline[language=Python]{NetLogoControllerServer}, which is responsible for starting and stopping the NetLogo model, sending commands to the model, fetching results from reporters on the model, querying parameters, and scheduling reporters over simulation execution. Each \lstinline[language=Python]{HeadlessWorkspaceController} object runs on a separate JVM thread. For batch simulation runs, NL4Py is able to relieve the Python application of thread/process management with \lstinline[language=Python]{run_experiment} by supplying a setup function and initialization data, allowing the \lstinline[language=Python]{NetLogoControllerServer} to handle thread and workspace management internally using JVM threads.

\subsection{Software Functionalities}
\label{sec:functionalities}

%Present the major functionalities of the software.
NL4Py can be imported and initialized with the path to the NetLogo directory using:
\begin{lstlisting}[language=Python]
import nl4py
nl4py.initialize(netlogo_home : str)
\end{lstlisting}

\lstinline[language=Python]{NetLogoHeadlessWorkspace}s can be created with the following function:
\begin{lstlisting}[language=Python]
nl4py.create_headless_workspace() -> nl4py.NetLogoHeadlessWorkspace.NetLogoHeadlessWorkspace
\end{lstlisting}

The NetLogo application can be started in GUI mode with the \lstinline[language=Python]{NetLogoGUI} object using:
\begin{lstlisting}[language=Python]
nl4py.netlogo_app() -> nl4py.NetLogoGUI.NetLogoGUI
\end{lstlisting}

The following functions can be then used to open and close NetLogo models on \lstinline[language=Python]{NetLogoHeadlessWorkspace}s (or \lstinline[language=Python]{NetLogoGUI}), respectively:
\begin{lstlisting}[language=Python]
nl4py.NetLogoHeadlessWorkspace.NetLogoHeadlessWorkspace.open_model(self, path : str)
nl4py.NetLogoHeadlessWorkspace.NetLogoHeadlessWorkspace.close_model(self)
\end{lstlisting}

The \lstinline[language=Python]{NetLogoHeadlessWorkspace} and \lstinline[language=Python]{NetLogoGUI} objects can execute NetLogo commands and reporters with the \lstinline[language=Python]{command} and \lstinline[language=Python]{report} functions. The \lstinline[language=Python]{command} function takes NetLogo syntax as Python strings and executes the command on the respective workspace, while the \lstinline[language=Python]{report} function takes in NetLogo syntax as strings, executes the reporter, and returns the results from the simulation.
\begin{lstlisting}[language=Python]
nl4py.NetLogoHeadlessWorkspace.NetLogoHeadlessWorkspace.command(self, command : str)
nl4py.NetLogoHeadlessWorkspace.NetLogoHeadlessWorkspace.report(self, reporter : str) -> Any
\end{lstlisting}

Reporters can be scheduled with the \lstinline[language=Python]{schedule_reporters} function. A list of NetLogo \lstinline[language=Python]{reporters} as Python strings is required. Additionally, the tick to \lstinline[language=Python]{start} and \lstinline[language=Python]{stop} reporting, and tick \lstinline[language=Python]{interval} between reporter execution can be provided. Optionally, a custom \lstinline[language=Python]{go_command} can be supplied, besides the default `go'. The results are returned as a dictionary (keys are ticks) of dictionaries (keys are reporters and values are reporter results).
\begin{lstlisting}[language=Python]
nl4py.NetLogoHeadlessWorkspace.NetLogoHeadlessWorkspace.schedule_reporters(self, reporters : list, startAtTick : int = 0, intervalTicks : int = 1, stopAtTick : int = -1, goCommand : str = 'go') -> List[List[Any]
\end{lstlisting}

The following functions can be used to query a NetLogo model's parameter names and the default ranges, and set the model parameters to random values, with the following functions, respectively:
\begin{lstlisting}[language=Python]
nl4py.NetLogoHeadlessWorkspace.NetLogoHeadlessWorkspace.get_param_names() -> List[str]
nl4py.NetLogoHeadlessWorkspace.NetLogoHeadlessWorkspace.get_param_ranges() -> List[Any]
nl4py.NetLogoHeadlessWorkspace.NetLogoHeadlessWorkspace.set_params_random()
\end{lstlisting}

\lstinline[language=Python]{NetLogoHeadlessWorkspace}s can be deleted with the following command:
\begin{lstlisting}[language=Python]
nl4py.delete_headless_workspace(headlessWorkspace : nl4py.NetLogoHeadlessWorkspace)
\end{lstlisting}

Finally, a batch of NetLogo runs may be executed with \lstinline[language=Python]{run_experiment}. This function requires, a \lstinline[language=Python]{callback} function that must return a list of NetLogo setup commands to initialize each simulation instance and a list of input \lstinline[language=Python]{data} to be passed to the callback function per simulation. Additionally, a list of \lstinline[language=Python]{reporters} is required, and the ticks to \lstinline[language=Python]{start} and \lstinline[language=Python]{stop} reporting, tick \lstinline[language=Python]{interval} between reporter execution, and simulation start \lstinline[language=Python]{go_command} can be supplied. The \lstinline[language=Python]{callback} function must accept elements from \lstinline[language=Python]{data} as a parameter. A simulation is started for each element in \lstinline[language=Python]{data}, by passing the element into \lstinline[language=Python]{callback} and initializing each simulation by executing the list of NetLogo commands returned by executing \lstinline[language=Python]{callback}. Each simulation is executed on a JVM worker thread and the number cores to be used can be optionally provided, otherwise all available cores are used by default. Each worker thread is assigned a chunk of the resulting list of lists of setup commands from executing \lstinline[language=Python]{callback} on \lstinline[language=Python]{data} and the results are returned as a Pandas \lstinline[language=Python]{DataFrame}.

\begin{lstlisting}[language=Python]
nl4py.run_experiment(model_name : str, callback : Callable[[Any], List[str]], data : List[Any], reporters : List[str], start_at_tick : int = 0, interval : int = 1, stop_at_tick : int = 1000000, go_command : str = "go", num_procs : int = multiprocessing.cpu_count()) -> pandas.DataFrame
\end{lstlisting}

\subsection{Performance Benchmarking}
\label{sec:analysis}
We benchmarked the performance of running parallel NetLogo simulations, on local and cloud environments with NL4Py, against PyNetLogo \cite{jaxa2018pynetlogo} and NetLogo's BehaviorSpace tool. PyNetLogo is an alternative solution to controlling NetLogo models through Python that uses the Java Native Interface, instead of socket communication. BehaviorSpace is NetLogo's GUI-based tool, implemented in Java and Scala, for performing batch simulation runs for factorial experiments, and does not provide the degree of programmatic simulation control offered by either NL4Py or PyNetLogo via the NetLogo controlling API. BehaviorSpace can also be executed via command-line and command-line execution was used for performance comparisons below.

We compared simulations of the Fire \cite{wilenskyfire}, Ethnocentrism \cite{wilenskyethnocentrism,axelrod2003evolution}, and Wolf Sheep Predation \cite{wilensky1997netlogo} NetLogo sample models. 
\lstinline[language=Python]{schedule_reporters} and \lstinline[language=Python]{run_experiment} of NL4Py were compared against \lstinline[language=Python]{repeat_report} of PyNetLogo and BehaviorSpace command-line execution. 
\lstinline[language=Python]{schedule_reporters} and \lstinline[language=Python]{repeat_report} used Python-based parallelism through the multiprocessing library, while \lstinline[language=Python]{run_experiment} uses JVM threads on the Java-based \lstinline[language=Python]{NetLogoControllerServer} similar to BehaviorSpace. 
The number of model runs were varied from 200 to 1000 for each model, and simulations were executed for 100, 1000, and 100 ticks for the Fire, Ethnocentrism, and Wolf Sheep Predation models, respectively. 
All three configurations were compared on both a Intel(R) Corei7-7700  CPU  (7th  generation) desktop PC with  16GB  of  RAM  running  Windows  10  (64Bit), and an AWS m5.8xlarge EC2 instance with 32 cores and 128 GB memory. 
For each configuration, the number of NetLogo workspaces initialized and executed in parallel was equal to the number of cores on the hardware being used. 

\begin{figure}[h!]%{width=1\linewidth}
    \centering
    \includegraphics[width= \linewidth]{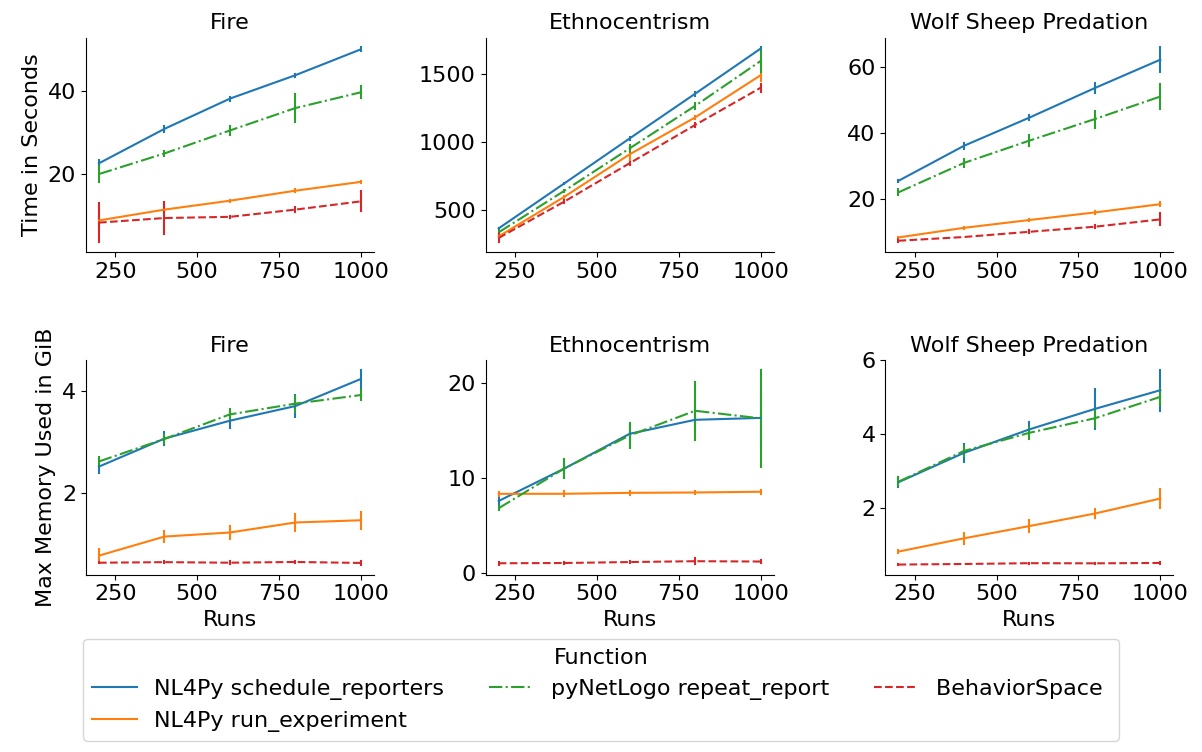}
    \caption{Execution time and maximum memory usage comparisons between NL4Py parallelized  \lstinline[language=Python]{scheduled_reporters}, NL4Py  \lstinline[language=Python]{run_experiment}, and PyNetLogo \lstinline[language=Python]{repeat_report} over varying model runs on a 8 core Desktop machine running Windows 10.}
    \label{fig:5_2_local}
\end{figure}

Fig. \ref{fig:5_2_local} compares the mean execution times and maximum memory usage (and 95\% confidence intervals of means) reported on the Windows desktop machine for the three different models over increasing run counts. Execution time is significantly lowered when using NL4Py's \lstinline[language=Python]{run_experiment} over the other two Python functions, nearing the execution time of BehaviorSpace, for all three models. Maximum memory usage is also higher for both PyNetLogo's \lstinline[language=Python]{repeat_report} and NL4Py's \lstinline[language=Python]{scheduled_reporters} in comparison to NL4Py's \lstinline[language=Python]{run_experiment}, which has roughly half of the memory footprint of the former two functions in most cases. In comparison, BehaviorSpace has a much smaller memory usage than all three Python functions. 

\begin{figure}[h!]%{width=1\linewidth}
    \centering
	\includegraphics[width= \linewidth]{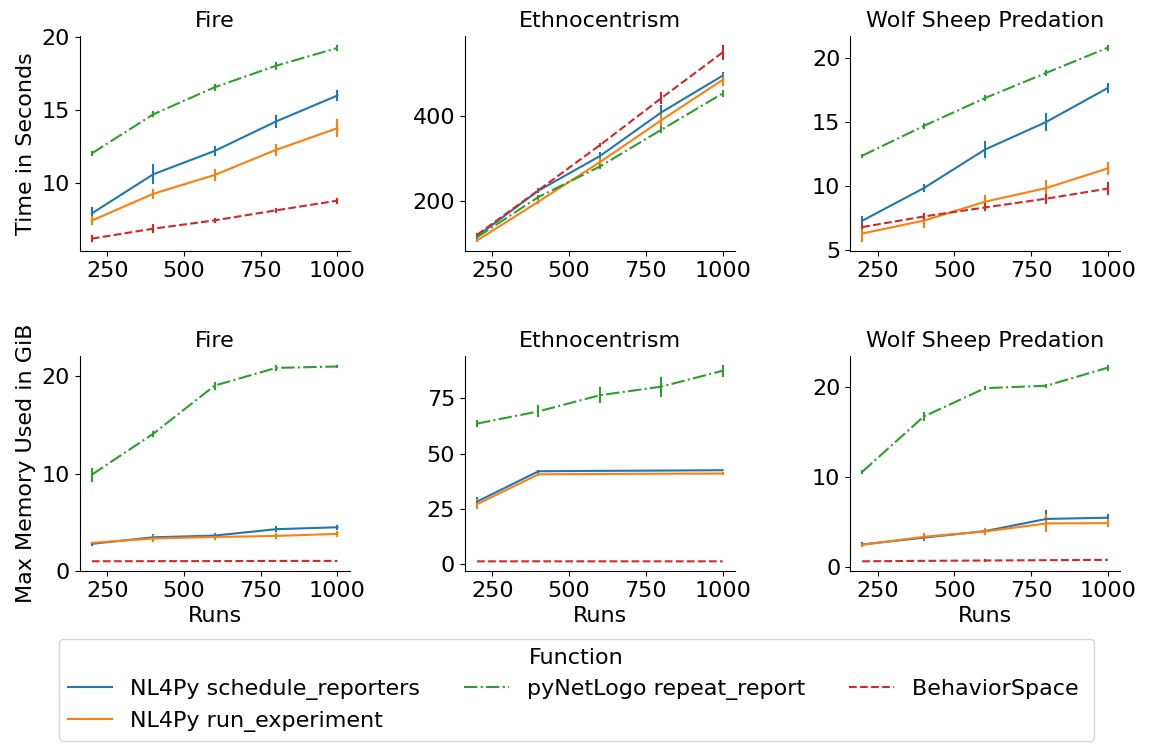}
    \caption{Execution time and maximum memory usage comparisons between NL4Py parallelized  \lstinline[language=Python]{scheduled_reporters}, NL4Py  \lstinline[language=Python]{run_experiment}, and PyNetLogo  \lstinline[language=Python]{repeat_report}, over varying model runs on an AWS m5.8xlarge EC2 instance (32 cores).}
    \label{fig:5_2_aws}
\end{figure}

Fig. \ref{fig:5_2_aws} compares the mean execution times and maximum memory usage (and 95\% confidence intervals of means) on the EC2 instance. Again, NL4Py's \lstinline[language=Python]{run_experiment} has signitificantly lower execution times than \lstinline[language=Python]{repeat_report} and \lstinline[language=Python]{schedule_reporters}, for the Fire and Wolf Sheep Predation models, with similar execution times to BehaviorSpace on Wolf Sheep Predation. However, for the Ethnocentrism model all three Python functions have lower execution times than BehaviorSpace for simulations with more than 600 runs. \lstinline[language=Python]{run_experiment} and \lstinline[language=Python]{schedule_reporters} have very similar memory usage for all three models, while memory usage is consistently very high for PyNetLogo's \lstinline[language=Python]{repeat_report} (over 4 times more for 1000 runs of Fire and Wolf Sheep Predation, and 2 times more for 1000 runs of Ethnocentrism). Again BehaviorSpace has a much lower memory usage than all three Python functions. 

Overall, it is seen that NL4Py's \lstinline[language=Python]{run_experiment} is the fastest Python function and is capable of reasonably nearing the execution times of BehaviorSpace compared to the other two Python functions. This can be attributed to its hybrid approach of using JVM threads for parallelization, similar to BehaviorSpace. All three Python methods generally use more memory than BehaviorSpace, due to the need to maintain additionally objects in memory to communicate with the NetLogo controlling API, while NL4Py demonstrates lower memory usage than PyNetLogo in many cases, especially on the cloud environment.

\section{Illustrative Examples}
\label{sec:examples}

%Provide at least one illustrative example to demonstrate the major functions.

%Optional: you may include one explanatory video that will appear next to your article, in the right hand side panel. (Please upload any video as a single supplementary file with your article. Only one MP4 formatted, with 50MB maximum size, video is possible per article. Recommended video dimensions are 640 x 480 at a maximum of 30 frames/second. Prior to submission please test and validate your .mp4 file at $ http://elsevier-apps.sciverse.com/GadgetVideoPodcastPlayerWeb/verification$. This tool will display your video exactly in the same way as it will appear on ScienceDirect.).

\subsection{Factorial Experiment}
\label{sec:parameters}
\begin{figure}[htbp]
  	\centering
	\includegraphics[width=\linewidth]{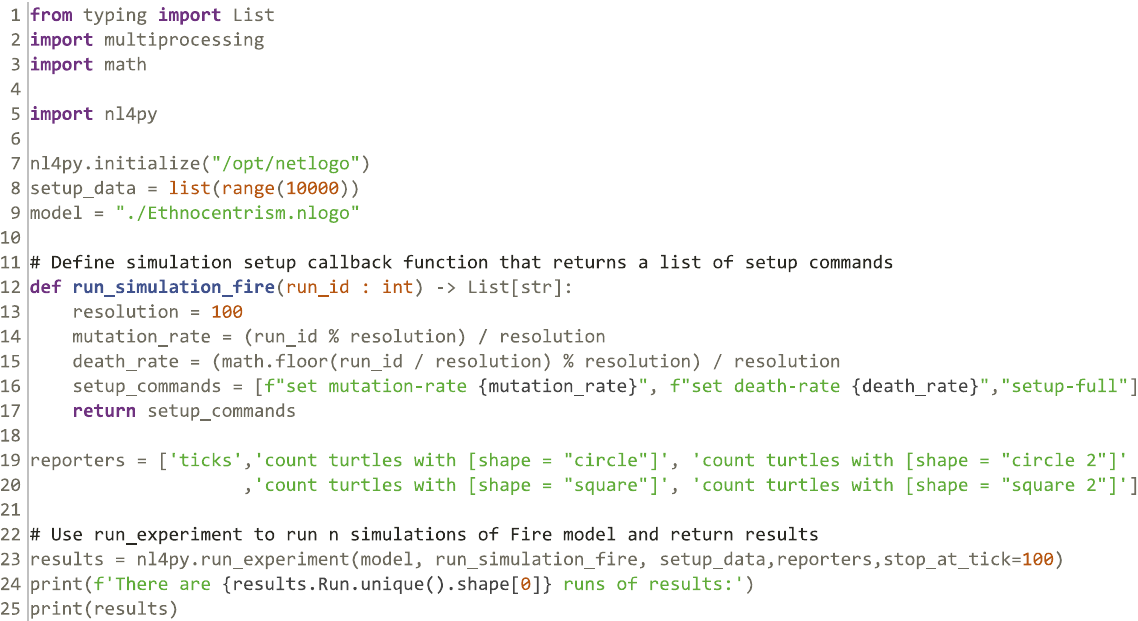}
	\caption{Example Python code for performing a parallelized factorial experiment on the Ethnocentrism model  with NL4Py's \lstinline[language=Python]{run_experiment}.}
    	\label{fig:paramcode}
\end{figure}

A batch of NetLogo simulations can be run using \lstinline[language=Python]{run_experiment}, which automatically handles parallelization. The example code shown in Fig. \ref{fig:paramcode} demonstrates how a parallelized factorial experiment on the Fire.nlogo sample model with 100 runs each of 100 possible parameter values, for a total of 10000 simulation runs, can be set up with only a few lines of code with NL4Py.

\subsection{Calibration}
\label{sec:calibration}
Calibration is a commonly used technique in agent-based modeling for finding the optimal parameter configurations required to produce a particular target outcome. Calibration has become such an integral part of the methodology that agent-based modeling platforms have been accompanied with ready-made calibration tools, such as OptQuest for AnyLogic and BehaviorSearch for NetLogo. However, if the user desires to use an optimization algorithm outside of those implemented by the provided calibration tool, they would require a way to control the agent-based model from their own code. 

\begin{figure}[htbp]
  	\centering
    \caption{Python code for model calibration with NL4Py and DEAP. NL4Py is used for extracting model parameter names and ranges, defining functions for initializing, executing, and measuring NetLogo simulations by the genetic algorithm individuals.}
    \label{fig:calibcode}
\includegraphics[width= 0.75\linewidth]{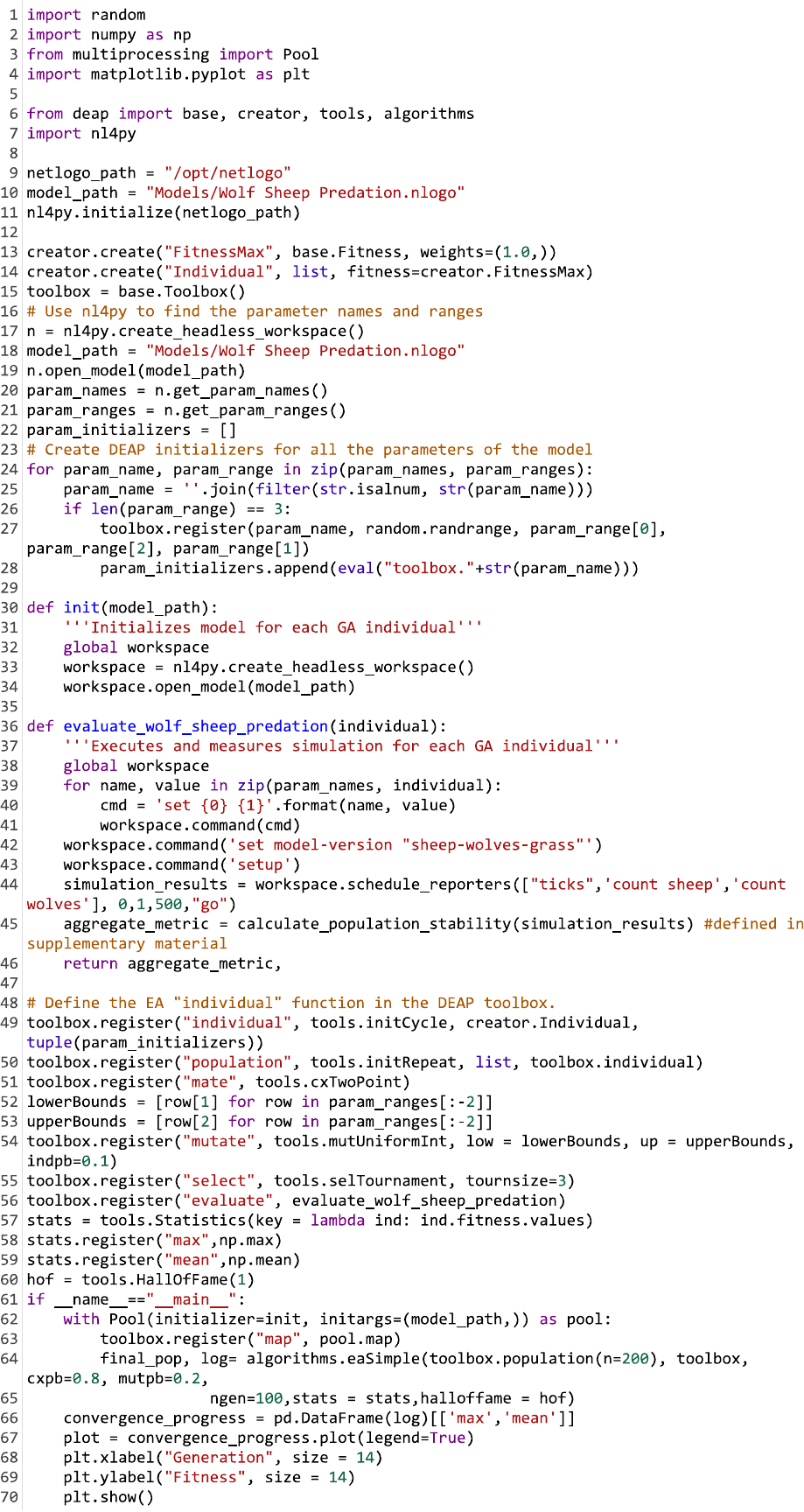}
\end{figure}

\begin{figure}[htbp]
  	\centering
    \caption{Convergence of the evolutionary algorithm towards optimal parameters for the Wolf Sheep Predation Model.}
    \label{fig:calibresults}
\includegraphics[width= \linewidth]{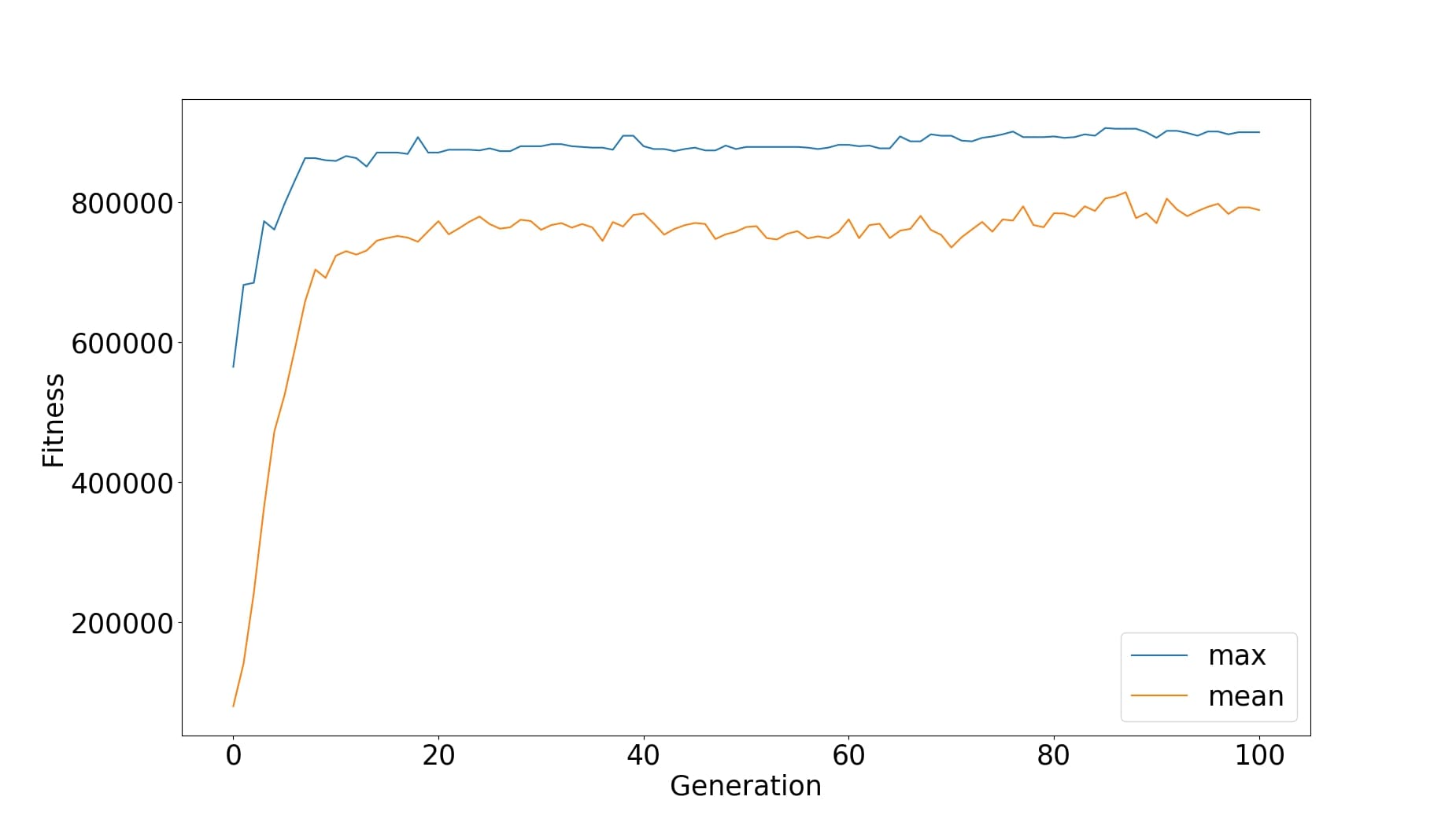}
\end{figure}

The Python package index has several optimization packages, from which we choose DEAP \cite{DEAP_JMLR2012}, due to its popularity, ease of use, and parallelization support. We demonstrate how DEAP and NL4Py can be used to configure and run an evolutionary algorithm (EA) to calibrate the Wolf Sheep Predation model in Fig. \ref{fig:calibcode}.
The objective of the calibration process in this experiment was finding a parameter configuration that was able to acheive a dynamic equilbrium of both wolf and sheep populations defined by the function \lstinline[language=Python]{calculate_population_stability}. This fitness function is described fully in \ref{app:pop}. Fig. \ref{fig:calibresults} displays the gradual convergence of the EA towards a solution. The maximum and mean fitness of the EA individuals increase rapidly within the first 20 generations after which fitter individuals are only found at a slower rate.

\section{Impact}
\label{sec:impact}

%\textbf{This is the main section of the article and the reviewers weight the description here appropriately}

%Indicate in what way new research questions can be pursued as a result of the software (if any).
Python has become the language of preference for most scientific computing and data analytics research, while NetLogo remains the software used for most agent-based modeling research. NL4Py enables researchers looking to scrutinize NetLogo models with powerful machine learning and analytical libraries that are freely available in Python. The complexity inherent to complex adaptive systems modeled using NetLogo causes even the simplest model to produce a vast landscape of possible outcomes. In response, NL4Py allows researchers to easily parallelize their experiments in Python to utilize high-performance computing infrastructure and perform the massive number of simulations required within reasonable time.

%Indicate in what way, and to what extent, the pursuit of existing research questions is improved (if so).
Agent-based models are often subjected to factorial experiments (section \ref{sec:parameters}) and other sensitivty analysis techniques, and parameter calibration (section \ref{sec:calibration}). These experiments often require vast numbers of simulation runs corresponding to the number of permuations of model parameter ranges or the number of iterations an optimization algorithm needs for model calibration. Due to their inherent stochasticity, agent-based models often require multiple simulations of each parameter configuration for a statistically valid description of their outputs. We have provided example scripts for the convenience of users looking to quickly set up such experiments with NetLogo models using NL4Py.

%Indicate in what way the software has changed the daily practice of its users (if so).
NL4Py allows researchers using NetLogo as their agent-based modeling framework to move beyond the optimization algorithms implemented in BehaviorSearch for parameter calibration (random search, simulated anealing, gradient descent, and genetic algorithm) and design more sophisticated experiments of their choice. Python has a rich ecosystem of machine learning and data analysis algorithms  that researchers can now utilize to assemble easily extensible scripts that perform experiments such as sensitivity analysis, without the burden of setting up a compilable code infrastructure in Java or Scala that can work with the NetLogo controlling API.

%Indicate how widespread the use of the software is within and outside the intended user group.
NL4Py has enabled several research projects that required external manipulation of NetLogo models additional to the evolutionary model discovery project it was initially designed for. Specifically, NL4Py has enabled sensitivity analysis and feature engineering for models of bacterial grown\footnote{pending publication}, agent-based modeling of emergency evacuation\cite {vandewalle2019integrating}, modeling production scheduling \cite{elmenreich2021artificial}, and the prediction of tipping points in complex social systems \cite{fullsack2020predicting}. These projects are examples where domain scientists, in this case, micro-biologists, operations researchers, industrial engineers, and computational social scientists, benefited from the parallelized, external control of NetLogo models that NL4Py offers. Furthermore, the GitHub project has garnered 27 stars and 4 forks at the time of writing, and users have brought up several issues that have led to the improvement of the software over the course of time. 

%Indicate in what way the software is used in commercial settings and/or how it led to the creation of spin-off companies (if so).

\section{Conclusions}
\label{sec:conclusions}

%Set out the conclusion of this original software publication.
We developed NL4Py as a means through which the agent-based modeling community can externally control and measure simulations of NetLogo models through Python in an easily parallelizable fashion. With NL4Py, developers can now open, set parameters, send commands, and schedule reporters to NetLogo models. Importantly, this enables machine learning and data analysis experiments to be performed on NetLogo models, which may have vast fitness landscapes and require many parallel simulations to explore. NL4Py is platform independent, compatible with Python 3.6+ and has been tested to work on Windows 10, MacOS 10, and Ubuntu operating systems.
NL4Py's client-server architecture demonstrates reasonable performance improvements in memory usage and execution time, under the differing conditions of both desktop environments and virtual cloud instances. We demonstrate how calibration can be performed on a NetLogo sample model using NL4Py and DEAP. We also provide an example on how a complete batch of parallelized NetLogo simulations can be easily executed and reportered on, with a few lines of code by allowing NL4Py to manage the details of parallelization. 
Our hope is that this contribution will increase the accessibility of NetLogo to computational researchers and domain scientists. We see NL4Py as a vital link, allowing the agent-based modeling community to utilize the latest technological advancements in data analytics librariess that are increasingly available on the Python package index.

\section{Conflict of Interest}
%Please select the appropriate text:

%Potential conflict of interest exists:
%We wish to draw the attention of the Editor to the following facts, which may be considered as potential conflicts of interest, and to significant financial contributions to this work. The nature of potential conflict of interest is described below: [Describe conflict of interest]

%No conflict of interest exists:
We wish to confirm that there are no known conflicts of interest associated with this publication and there has been no significant financial support for this work that could have influenced its outcome.

\section*{Acknowledgements}
\label{}

%Optionally thank people and institutes you need to acknowledge. 
This work was supported by DARPA program HR001117S0018 (FA8650-18-C-7823).
%% The Appendices part is started with the command \appendix;
%% appendix sections are then done as normal sections
%% \appendix

%% \section{}
%% \label{}
\appendix
\section{Population Stability calculation for the Wolf Sheep Predation Model}
\label{app:pop}

The Wolf Sheep Predation model \cite{wilensky1997netlogo} is an agent-based model of a predator-prey system inspired by the Lotka-Volterra equations \cite{lotka1926,volterra1926fluctuations}. A NetLogo implementation of this model is provided with the NetLogo sample models suite. In the calibration experiment, we use NL4Py to control simulations of this NetLogo model and measure population stability. The objective of the experiment is to find the optimal parameters that can maximize a population stability score. Total population stability, $E_t$, was calculated at every simulation time step, $t$. $E_t$ was found as follows. We measured the first order derivatives of both the wolf population ($P_W$) and sheep population ($P_S$) over time, $\frac{dP_{W,t}}{dT}$ and $\frac{dP_{S,t}}{dT}$, by finding the change in population between each simulation tick (see Eqs. \ref{eq:ratechange} and \ref{eq:ratechangesimplified}). These derivatives were then passed through a unit step function, in order to score extinction of either species as $0$. The reciprocal of this result was considered the population stability score for each species at $t$, $E_{W,t}$ and $E_{S,t}$ (see Eq. \ref{eq:populationstabilityspecies}). We then took the mean population stability for both species, for each simulation step, to get a total population stability over time. Finally, an aggregate mean total population stability score, $score$, was found by dividing the result by the number of time steps simulated, $k$ (see Eq. \ref{eq:score}).  

\begin{equation} \label{eq:ratechange}
\frac{dP_{W,t}}{dT} = ((P_{W,t} - P_{W,t - 1}) / \Delta T  \quad and \quad
\frac{dP_{S,t}}{dT} = ((P_{S,t} - P_{S,t - 1}) / \Delta T  
\end{equation}
\textrm{$\Delta T = 1$, since the model runs in discrete simulation time units, or ticks. So,} 
\begin{equation} \label{eq:ratechangesimplified}
\frac{dP_{W,t}}{dT} = ((P_{W,t} - P_{W,t - 1})  \quad and \quad
\frac{dP_{S,t}}{dT} = ((P_{S,t} - P_{S,t - 1})
\end{equation}

\begin{equation} \label{eq:populationstabilityspecies}
E_{W,t} = 
\begin{dcases}
   \frac{1}{dP_{W,t}}, & \text{if } P_{W,t}\geq 1\\
      0,              & \text{otherwise}
\end{dcases} \quad and \quad
E_{S,t} = 
\begin{dcases}
   \frac{1}{dP_{S,t}},& \text{if } P_{S,t}\geq 1\\
      0,              & \text{otherwise}
\end{dcases}
\end{equation}
\begin{equation} \label{eq:score}
score = \frac{\sum_{t = 0}^{k} (E_{S,t} + E_{W,t})}{k}
\end{equation}

The Python function \lstinline[language=Python]{} is provided in Fig. \ref{}. 

\begin{figure}
  	\centering
	\includegraphics[width=\linewidth]{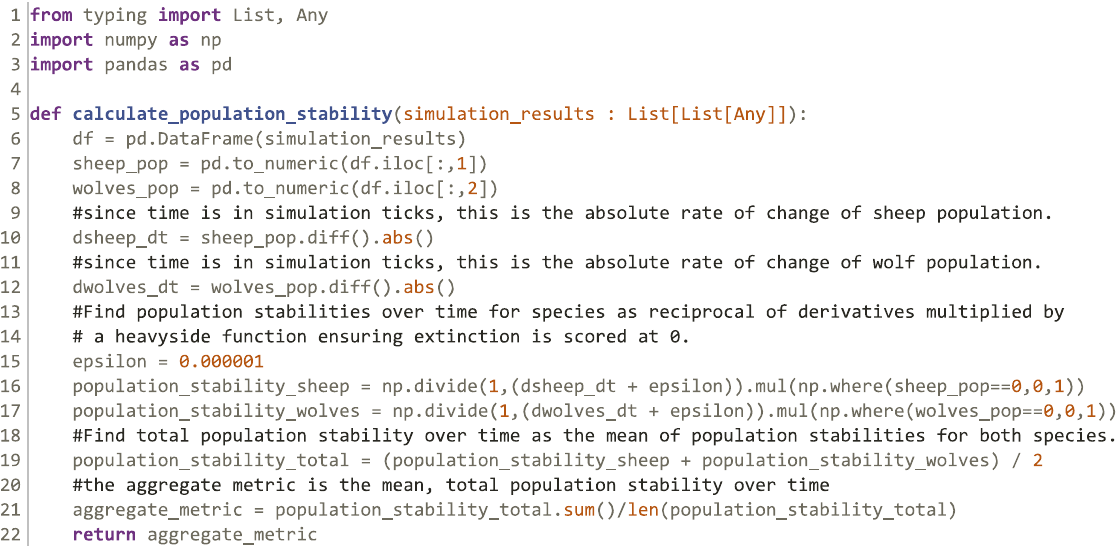}
	\label{fig:pop_stability_code}
	\caption{Python code for calculating population stability in the Wolf Sheep Predation model}
\end{figure}

%% References:
%% If you have bibdatabase file and want bibtex to generate the
%% bibitems, please use
%%
%%  \bibliographystyle{elsarticle-num} 
%%  \bibliography{<your bibdatabase>}
\bibliographystyle{elsarticle-num} 
%\bibliography{NL4Py}

%% else use the following coding to input the bibitems directly in the
%% TeX file.

%\begin{thebibliography}{00}

%% \bibitem{label}
%% Text of bibliographic item

%\end{thebibliography}

Please add the reference to the software repository if DOI for software  is available. 

\section*{Current executable software version}
\label{}

%Ancillary data table required for sub version of the executable software: (x.1, x.2 etc.) kindly replace examples in right column with the correct information about your executables, and leave the left column as it is.

\begin{table}[!h]
\begin{tabular}{|l|p{6.5cm}|p{6.5cm}|}
\hline
\textbf{Nr.} & \textbf{(Executable) software metadata description} & \textbf{Please fill in this column} \\
\hline
S1 & Current software version & 0.9.0 \\
\hline
S2 & Permanent link to executables of this version  & \url{https://pypi.org/project/NL4Py/} \\
\hline
S3 & Legal Software License & GNU General Public License 3.0 (GPLv3) \\
\hline
S4 & Computing platforms/Operating Systems & Linux, OS X, and Microsoft Windows 10 \\
\hline
S5 & Installation requirements \& dependencies & NetLogo, Java, Python 3.6+\\
\hline
S6 & If available, link to user manual - if formally published include a reference to the publication in the reference list & \url{https://github.com/chathika/NL4Py/blob/master/Readme.md} \\
\hline
S7 & Support email for questions & chathika@knights.ucf.edu \\
\hline
\end{tabular}
%\caption{Software metadata (optional)}
\caption{Software metadata}
\label{} 
\end{table}

\end{document}